\documentclass[espcrc2]{elsart}
\usepackage{amssymb,epsfig}
\begin{document}
\begin{frontmatter}
\title{Review of Recent and Future Needs in \\ Hadronic Flavor Particle Production Measurements}
\author{Dr. Nickolas Solomey}
\address{Illinois Institute of Technology, Chicago, USA and\\
Fermi Natinonal Accelerator Laboratory, Batavia, Ill. USA}
\begin{abstract}
Interactions of energetic particles on target nuclei producing secondary particles will be reviewed. Current 
simulation codes rely upon poorly measured results from the past. 
While current neutrino experiments, both atmospheric and accelerator based, rely upon Kaon and pion
production measurements which are poorly known and dominate their errors. The goal for the current round of experiments are to 
dramatically improve these measurements while improvments beyond this are still needed.
It is not only of interest to neutrino experiments, but also for designing calorimeters for the the International
Linear Collider which must achieve unprecedented resolutions for reaching their stated physics goals.
\end{abstract}
\end{frontmatter}

\section{Introduction}
\begin{figure}
\begin{center}
\mbox{
\epsfig{file=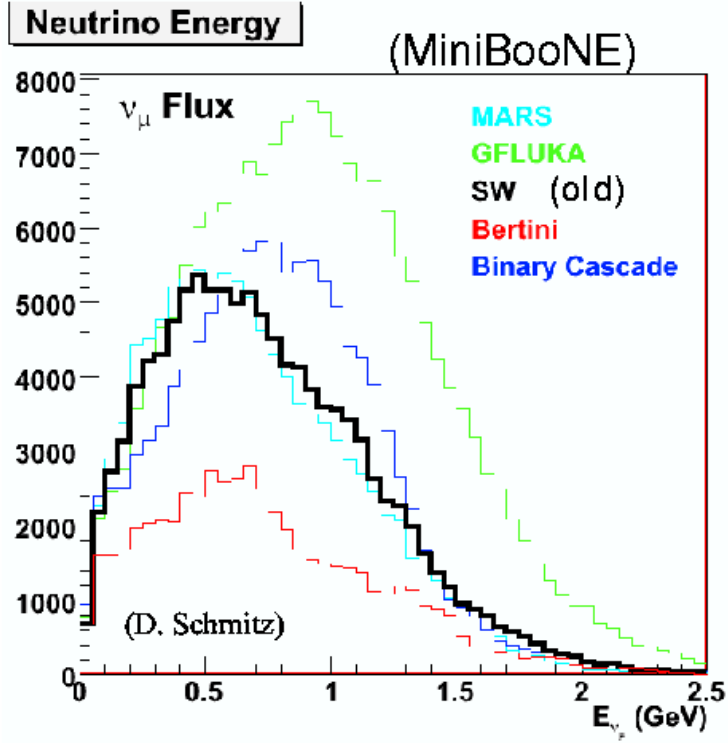,%
height=0.725\linewidth}
}
\end{center}
\caption{Calculated neutrino fluxes from a simulated MiniBooNE target using various codes and different
measurements.} \label{miniboon}
\end{figure}

Two new recently completed experiments studying hadronic flavor particle 
production, HARP-PS214 at CERN and MIPP-E907 at Fermilab have just completed 
data taking. The BNL E910 experiment which ended six years ago is still in the
data analysis process. Collectively these experiments cover 1 to 120 GeV/c on multiple targets 
(liquid Hydrogen, K2K and Minos targets and various nuclear targets 
including Uranium) for six beam species (pion, kaon, protons and their 
antiparticles). They expand upon the older cross section data and bring 
a variety of new physics goals: improved hadronic flavor production cross sections,
studying fundamental scaling law relationships \cite{raja} and nuclear target production studies.
Figure \ref{miniboon} shows the large uncertainty using current production cross-section data 
with a simulated neutrino target for the Fermilab MiniBooNE experiment \cite{schmitz}. This and other 
neutrino production experiments need substantial improvements, since this is what currently limits advances
in neutrino physics. 

\begin{figure}
\begin{center}
\mbox{
\epsfig{file=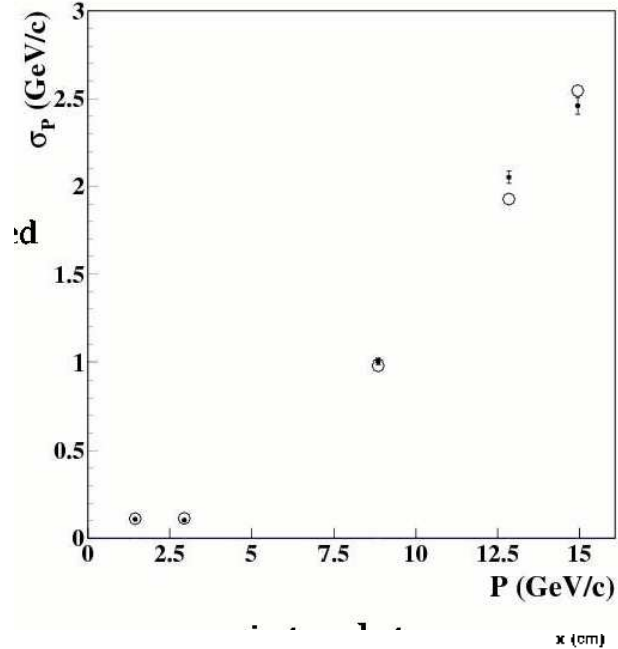,%
height=0.61\linewidth}
}
\mbox{
\epsfig{file=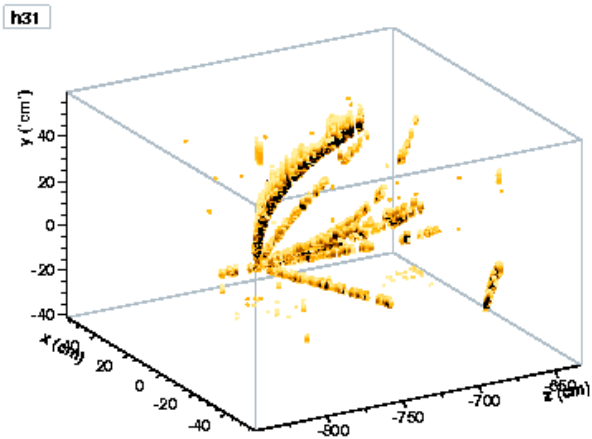,%
height=0.45\linewidth}
}
\end{center}
\caption{Momentum resolution from the Harp experiment (top) circle are MC and points for data, and an example 
event (bottom) from the MIPP experiment showing vertexing and high density tracking abilities of the TPC.} \label{tracking}
\end{figure}

\begin{figure}
\begin{center}
\mbox{
\epsfig{file=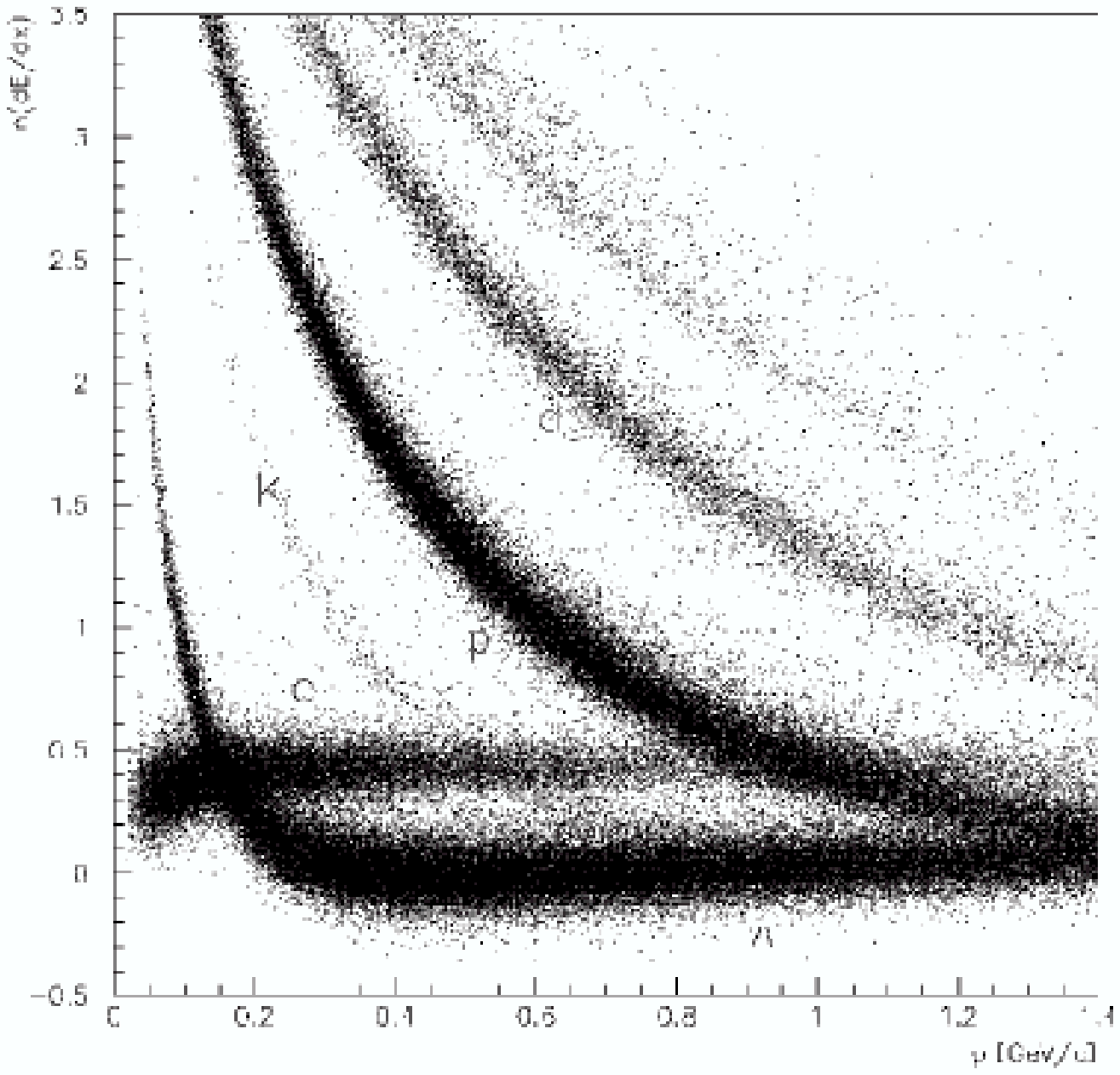,%
height=0.4\linewidth}
}
\mbox{
\epsfig{file=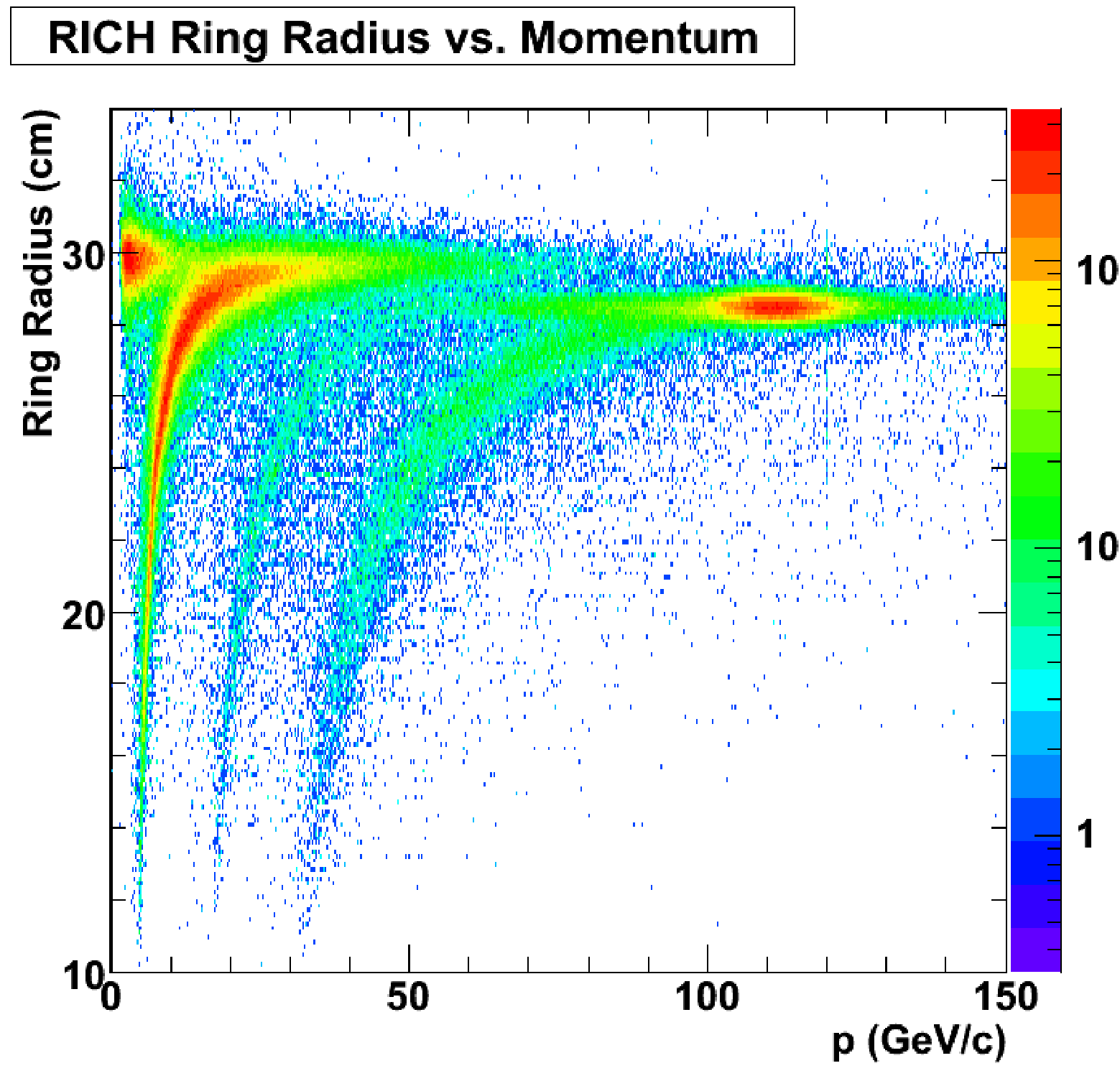,%
height=0.45\linewidth}
}
\end{center}
\caption{Two examples of the particle identification in these measurements are 
the E910 TPC using dE/dx on the left (this is the same TPC that
is currently in the Fermilab MIPP experiment) and on the right is the MIPP RICH detector.} \label{PID}
\end{figure}

In these new experiments particle tracking and identification are both paramount, bringing 
essential information, see figures \ref{tracking} and \ref{PID}. These new data set and the broad physics 
analysis planned with preliminary first results on all of these topics are 
presented in the next section. 

The proposed future run of an upgraded MIPP detector is of 
importance to future neutrino physics programs worldwide, such as 
Ice-Cube, NO$\nu$A, Pierre Auger and SuperK/HyperK experiments. 
By doing a high statistics run including Liquid Nitrogen, 
for the complete forward production hemisphere with 
three sigma particle identification, it will permit detailed cross section 
measurements to refine the Monte-Carlo generators such as Fluka or MARS, 
which are of major relevance for improving these future neutrino detectors. 
Current SuperK atmospheric neutrinos results are limited to a 20\% error \cite{suzuki} 
from these uncertainties while the Ice-Cube experiment sees a 30 to 40\% \cite{halzen} impact 
using the current known cross-sections.

\section{Recent results}
New results were reported by the Brookhaven National Laboratory E910 experiment within the last two years with 6.4 and 12.3 GeV/c 
protons on a Beryllium target \cite{e910}. 
The most recent published cross section results come from the Harp (CERN PS214) experiment. They reported
two measurements of 12.9 GeV/c proton on Aluminum \cite{harp1} of interest to the Japanese K2K experiment and 8.9 GeV/c protons 
on Beryllium \cite{harp2} which is of interest to the Fermilab MiniBooNE experiment.

\begin{figure}
\begin{center}
\mbox{
\epsfig{file=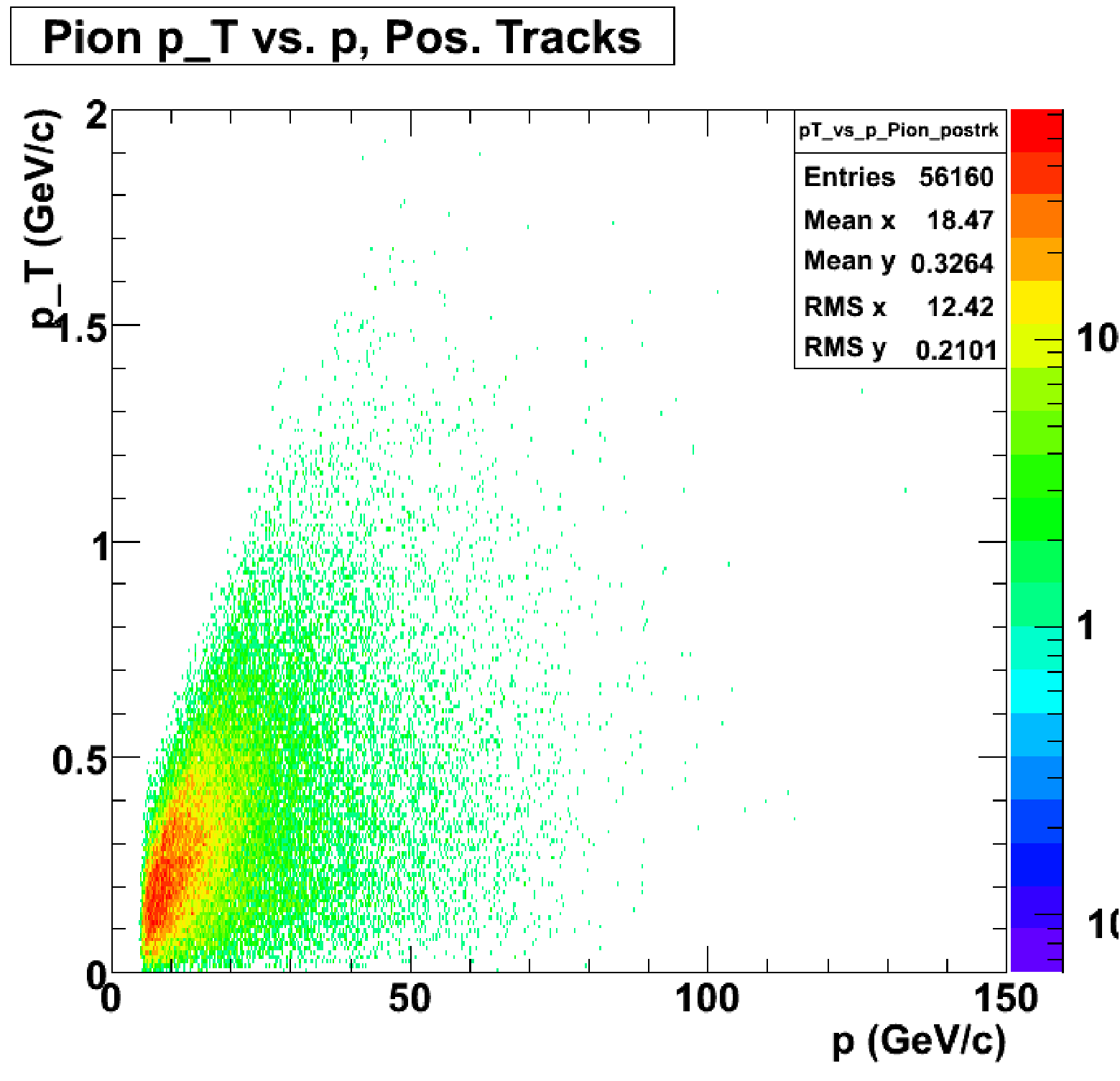,%
height=0.5\linewidth}
}
\mbox{
\epsfig{file=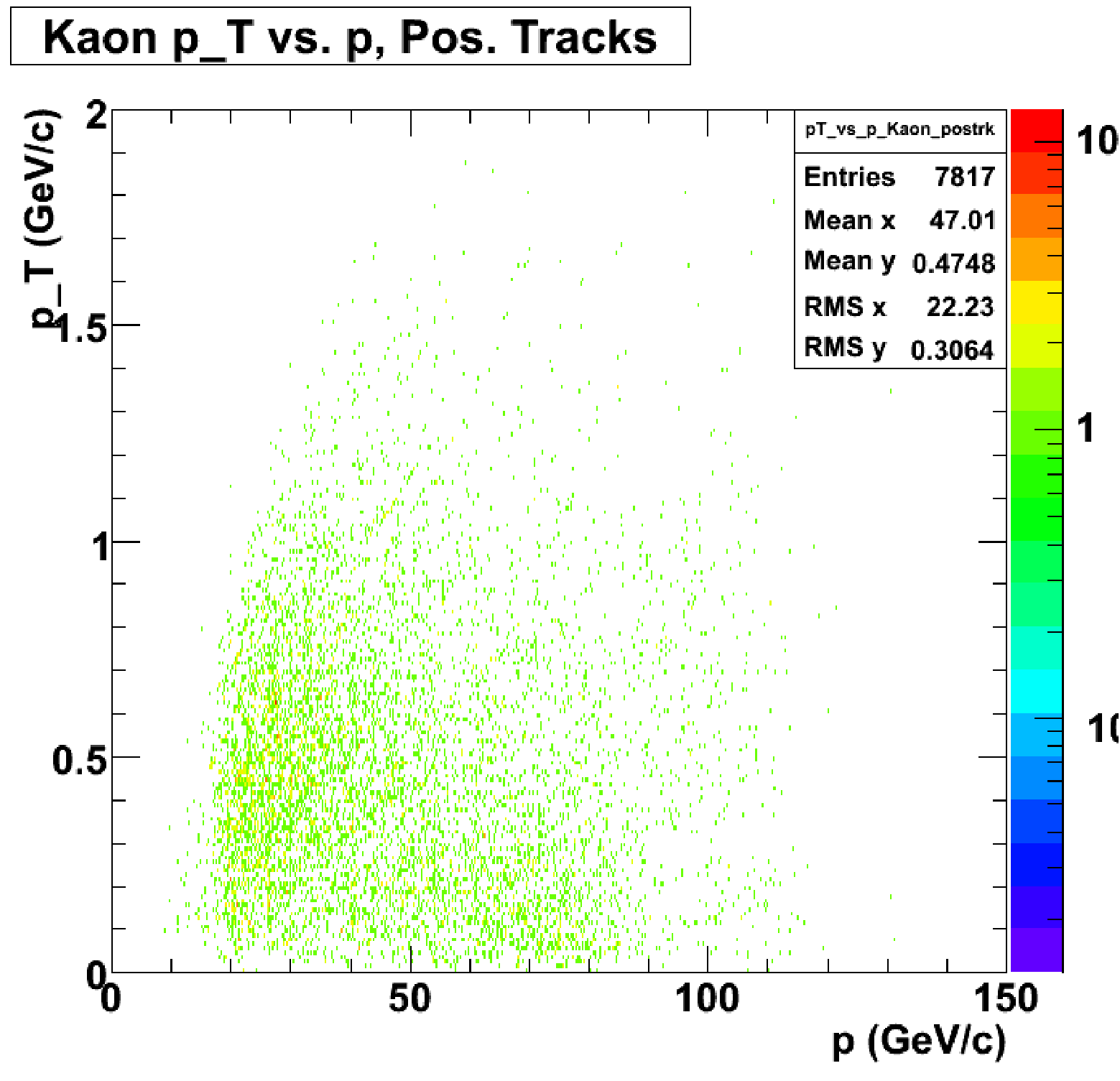,%
height=0.5\linewidth}
}
\end{center}
\caption{Preliminary study of $\pi$ (top) and $K$ (bottom) production by 120 GeV/c proton beam on the NuMI target 
in the MIPP experiment.} \label{mipp1}
\end{figure}

NuMI Target Studies for the MINOS Experiment with the MIPP data are well advanced.
Crucial $\pi$ and $K$ production studies that the MINOS analysis will eventually rely upon will be provided 
by the MIPP experiment using the spare NuMI composite target. A 120 GeV/c beam of pure 
protons identical to that which hits the NuMI horn target was run in the MIPP experiment for two 
months and collected 2 million events. These events using the projectile particle tracking and secondary particle 
production identification and momentum reconstruction show first results from this analysis in figure \ref{mipp1} \cite{paley}.

\begin{figure}
\begin{center}
\mbox{
\epsfig{file=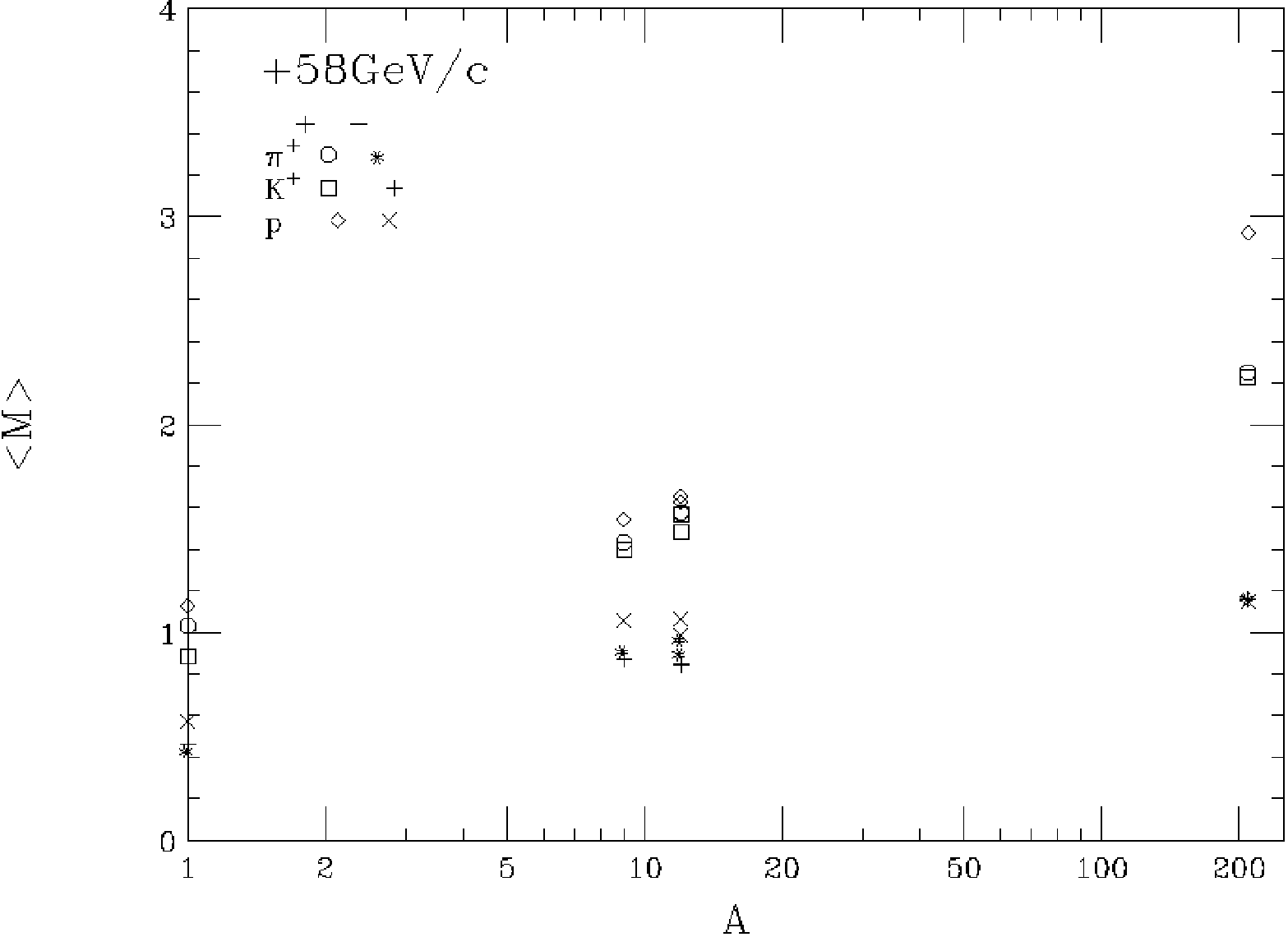,%
height=0.65\linewidth}
}
\end{center}
\caption{Preliminary MIPP results on the charged particle multiplicity from 0.15 to 1.0 GeV/c particles for beam particles at
+58 GeV/c with tagged $\pi$, K and protons on nuclear targets from A=1 hydrogen to
A=206 Bismuth, the data has been corrected for empty target runs.} \label{mipp2}
\end{figure}

The MIPP experiment is also studying particle production multiplicities and its A dependence, see figure \ref{mipp2},
for the first tentative results which are in agreement with \cite{liu}.

\section{Future improved measurements}
The Fermilab MIPP experiment proposes to have an upgraded run 
from 2007 to 2009 with a faster readout of the TPC so that a 3 kHz rate can be achieved, 4$\pi$ acceptance by installing
a backward hemisphere detector, a new silicon vertex-interaction trigger, an improved calorimeters which is a cooperation with the
International Linear Collider and other minor but essential modifications.

\begin{figure}
\begin{center}
\mbox{
\epsfig{file=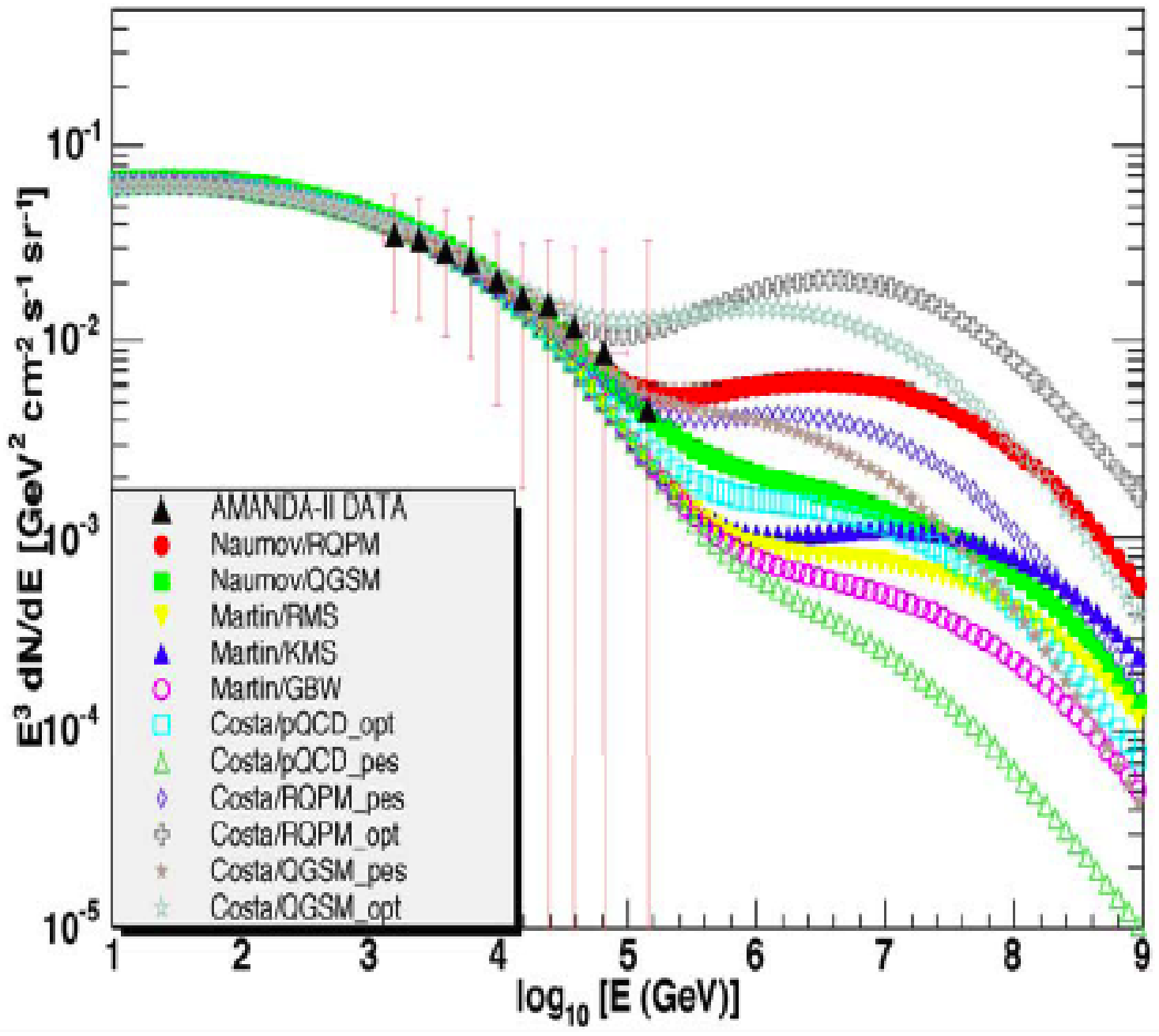,%
height=0.725\linewidth}
}
\end{center}
\caption{Errors in atmospheric neutrino production based upon current Nitrogen kaon production cross-section.} \label{amanda}
\end{figure}

\begin{figure}
\begin{center}
\mbox{
\epsfig{file=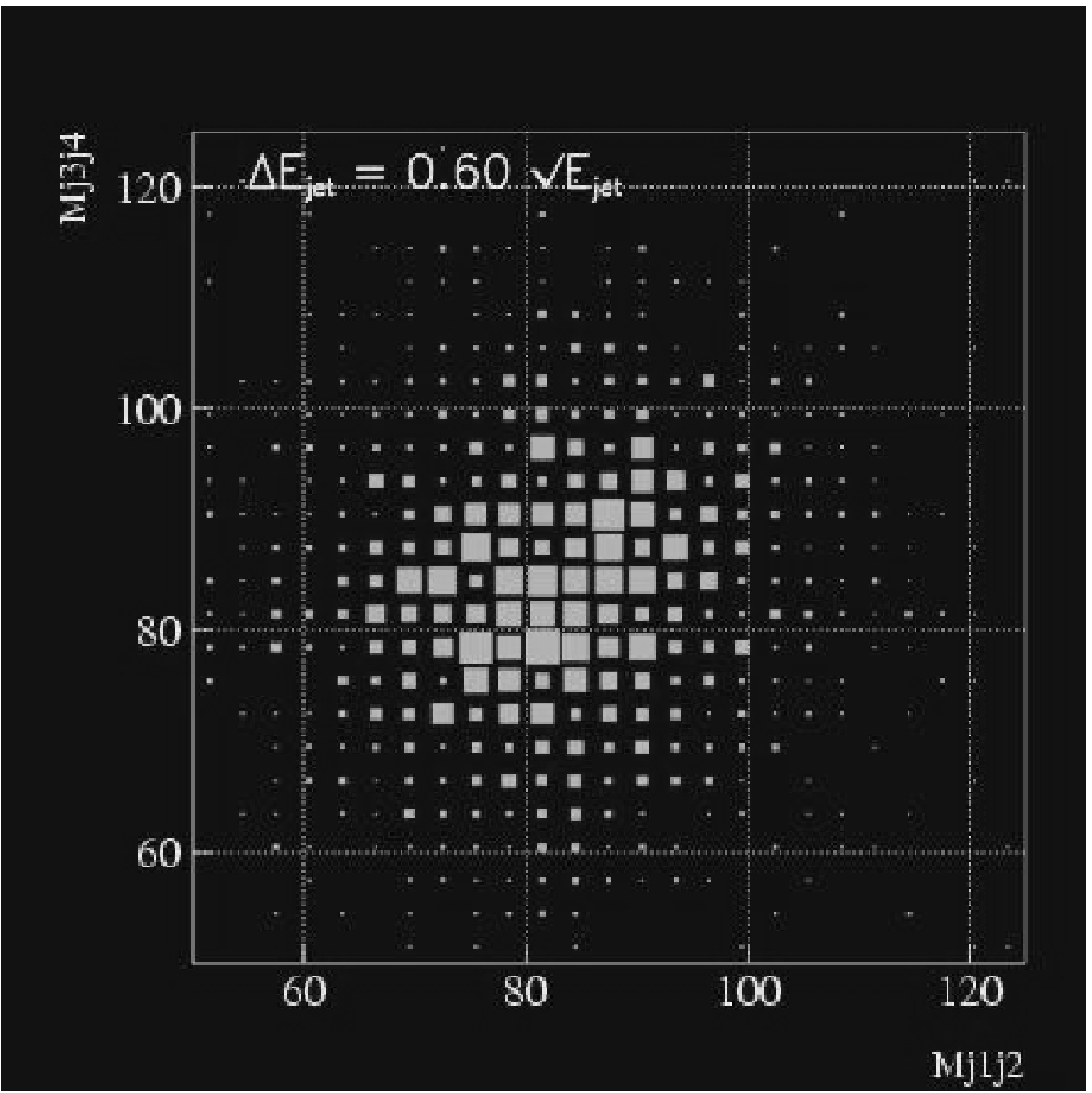,%
height=0.461\linewidth}
}
\mbox{
\epsfig{file=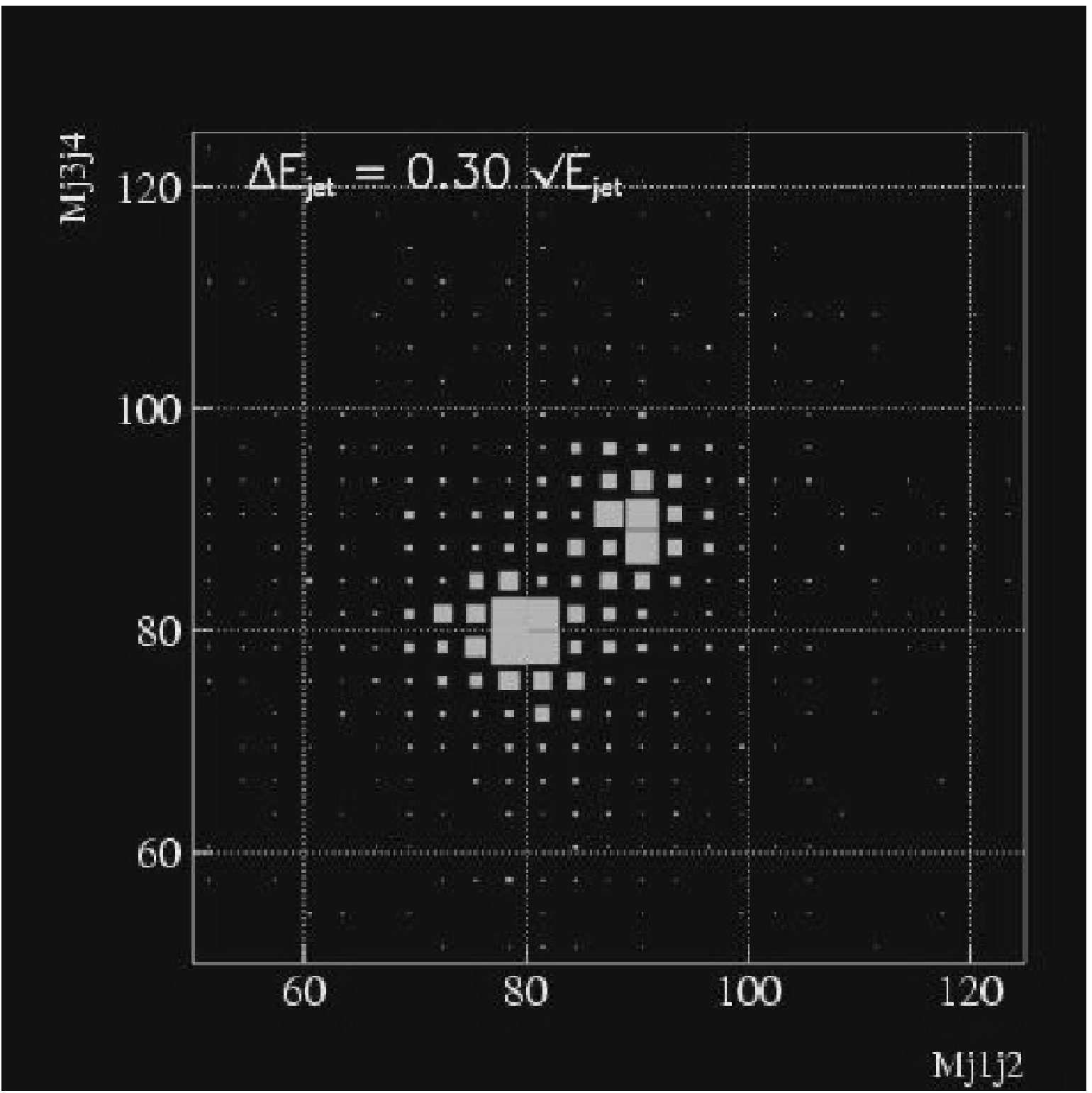,%
height=0.461\linewidth}
}
\end{center}
\caption{Hadronic calorimeter energy resolution at 60\%/$\sqrt{E}$ (left) and 30\%/$\sqrt{E}$ (right) for the sum of two jets on the x and y
axis.} \label{ILC_calo}
\end{figure}

An extensive run with a Liquid Nitrogen target with a large data 
sample so that fine segmentation in angle can be measured 
for all interaction species, this is explicitly needed for the atmospheric neutrino 
experiments such as: Ice-Cube and Hyper-K, to improve their neutrino flux from $\pi$ and 
K decays. Figure \ref{amanda} shows the current limitation to the small Amanda 
experiment above 4 GeV/c, this comes from the Cosmic Ray uncertainties of atmospheric production of 
Kaons and pions \cite{teresa}. Current models are uncertain above 5 GeV/c. Members of the Ice-Cube experiment from
the Univ. of Wisconsin and Fermilab members of the Pierre Auger Experiment will be joining the MIPP upgraded experiment.
It plans to study different elements for improvements to hadronic shower simulation 
code for hadronic calorimeter design.
These measurements are vital to improving Hadron Calorimetry, which is the most important 
detector in the ILC.  See figure \ref{ILC_calo} which shows the difference 
between a 60\%/$\sqrt{E}$ and 30\%/$\sqrt{E}$ 
resolution, a crucial necessity to separating the two jet background from W and Z. 

Hadron Calorimetry designs rely upon Monte-Carlo codes where the detector materials are not well 
known. The upgraded MIPP experiment will study 40 nuclei from 1 to 100 GeV/c.
The response of the ILC test calorimeter to neutrons and its efficiency is essential to 
the Particle Flow algorithms. By putting the ILC hadron calorimeter in the MIPP 
beam line they will be able to 
provide a direct measure of neutron energy and efficiency response with tagged 
neutrons \cite{n_tagging}. It will also provide a unique opportunity for tagged $K^0$ and $\bar{K}^0$ physics program. 

\section{Conclusion}
The E910, MIPP and Harp experiments with current improvements to the 
hadronic flavor production studies has greatly helped make improved cross-sections. Future 
data from the upgraded MIPP experiment will be a guiding light for both 
atmospheric and accelerator based neutrino experiments, without 
which there is little or no reason to continue neutrino studies. 
A great improvement in anti-proton interactions and charm production is also planned. 
Future data and analyzed results are excitingly awaited.

\end{document}